\documentclass[12pt,a4paper]{article}

\usepackage{jheppub} 
\usepackage[swedish,english]{babel}
\usepackage[utf8]{inputenc}
\usepackage[T1]{fontenc} 
\usepackage{graphicx}
\usepackage{amsmath}
\usepackage{amssymb}
\usepackage{amsthm}
\usepackage{latexsym}
\usepackage{color}
\usepackage{bm}
\usepackage{dsfont}
\usepackage{slashed}
\usepackage{hyperref}
\usepackage{breqn}
\usepackage{mathtools}
\usepackage{bm}
\usepackage{cleveref}
\usepackage[bottom]{footmisc}
\newcommand{\beq}{\begin{equation}}
\newcommand{\eeq}{\end{equation}}
\newcommand{\bea}{\begin{eqnarray}}
\newcommand{\eea}{\end{eqnarray}}

\def\beq{\begin{equation}}
\def\eeq{\end{equation}}
\def\bdm{\begin{displaymath}}
\def\edm{\end{displaymath}}

\def\bea{\begin{eqnarray}}
\def\eea{\end{eqnarray}}

\makeatletter
\newcommand{\pushright}[1]{\ifmeasuring@#1\else\omit\hfill$\displaystyle#1$\fi\ignorespaces}
\newcommand{\pushleft}[1]{\ifmeasuring@#1\else\omit$\displaystyle#1$\hfill\fi\ignorespaces}

\makeatother

\begin{document}

\title{Exotic Holographic Dispersion}


\author[a]{U.~Gran,}
\author[a]{M.~Torns\"o}
\author[b]{and T.~Zingg}


\affiliation[a]{Department of Physics,
Division for Theoretical Physics,
Chalmers University of Technology\\
SE-412 96 G\"{o}teborg,
Sweden}
\affiliation[b]{Nordita,
Stockholm University and KTH Royal Institute of Technology\\
Roslagstullsbacken 23,
SE-106 91 Stockholm,
Sweden}

\emailAdd{ulf.gran@chalmers.se}
\emailAdd{marcus.tornso@chalmers.se}
\emailAdd{zingg@nordita.org}

\abstract{For strongly interacting systems holographic duality is a powerful framework for computing e.g.~dispersion relations to all orders in perturbation theory. Using the standard Reissner--Nordstöm black hole as a holographic model for a (strange) metal, we obtain exotic dispersion relations for both plasmon modes and quasinormal modes for certain intermediate values of the charge of the black hole. 

The obtained dispersion relations show dissipative behavior which we compare to the generic expectations from the Caldeira--Leggett model for quantum dissipation. Based on these considerations, we investigate how holography can predict higher order corrections for strongly coupled physics.
}

\maketitle


\section{Introduction}
\label{sec:intro}


Analyzing strongly coupled systems is a challenging task since perturbative methods, and sometimes even numerical methods, are not readily applicable. During the past twenty years, however, the framework of holographic duality has been developed~\cite{Maldacena:1997re,Gubser:1998bc,Witten:1998qj}, in which a strongly coupled QFT is mapped to a dual weakly coupled gravitational system in one extra dimension, a system which can be analyzed using conventional techniques. The QFT is viewed as living on the boundary of the larger gravitational `bulk' space-time, hence the name `holographic' duality.
Holographic duality has been applied to wide ranging strongly coupled systems, e.g.~in string theory, QCD and condensed matter physics, with the hope of uncovering physical phenomena not accessible by standard techniques. In this paper we use holographic duality to compute dispersion relations which in a specific `intermediate' range of parameters, specified below, display exotic properties not captured by available weakly coupled models. 


We will consider the dispersion relations for both quasinormal modes (QNMs), corresponding to poles in `screened' density-density response,
\begin{equation}
    \chi_{sc} \;=\; \langle \rho\,\rho \rangle \,,
\end{equation}
and plasmon modes~\cite{Aronsson:2017dgf, Aronsson:2018yhi}, corresponding to poles in the `physical' density-density response $\chi$. For the longitudinal excitations we consider the `screened' and `physical' density-density response which are related via the longitudinal dielectric function  $\epsilon_{L}$ -- see e.g.~\cite{nozieres1966theory} for details -- through the simple relation
\begin{equation}
    \chi \;=\; \frac{\chi_{sc}}{\epsilon_{L}} \,.
\end{equation}
In the holographic dictionary, QNMs correspond to standard Dirichlet conditions for perturbations of bulk fields at infinity, while the plasmon modes can be identified with a specific type of mixed boundary conditions~\cite{Aronsson:2017dgf, Aronsson:2018yhi}. These mixed boundary conditions are related to a double trace deformation in the QFT~\cite{Witten:2001ua,Mueck:2002gm}, which can be argued to correspond to the form of the Green function within the random phase approximation (RPA)~\cite{Zaanen:2015oix}. This connection to conventional condensed matter formalism is an important piece of evidence in favour of the validity of the holographic treatment of plasmons~\cite{Aronsson:2017dgf, Aronsson:2018yhi}.


We compare the holographic dispersion relations we obtain to the generic expectation from the Caldeira--Leggett (CL) model of quantum dissipation~\cite{Caldeira:1982iu,Caldeira:1982uj} of a system coupled to its environment.
A key property of the CL model is that it is analytically tractable and thus provides an exact\footnote{Within the assumptions of the model, in particular that the system can be treated in the adiabatic approximation in relation to the environment.} quantum treatment of dissipation. Since only first order perturbation theory is used in the analysis of the CL model it is not expected to yield the correct result for strongly coupled systems and 
higher order corrections 
are
necessary to describe the holographic dispersion relations we obtain.


One of the generic results from the CL model is the appearance of a damping term due to dissipation. Qualitatively, this would lead to a toy model for a standard (weakly coupled) dispersion relation for a bulk plasmon\footnote{For surface plasmons, having a dispersion relation of the form $\omega(k) \propto \sqrt{k}$, we do not find any `exotic' dispersion relations using holography~\cite{Aronsson:2018yhi}.}
\begin{equation}
    \label{eq:bulkplasmon}
    \omega^2 - c^2 k^2 + i \omega \eta - \upsilon = 0 \,.
\end{equation}
The parameter $c$ is a speed of sound, $\upsilon$ is a potential term, usually including a mass term $m$, and $\eta$ is a damping parameter stemming from interaction with an underlying, second quantum system. Depending on the size of $\eta$ there are two possible and qualitatively different forms of the dispersion relation for small and intermediate momenta. If $\eta$ is small enough, expected for a weakly coupled system, the form of the standard quadratic dispersion would not change much. However, if damping $\eta$ exceeds a certain critical value the system becomes over-damped and the qualitative behavior changes -- we illustrate this in figure~\ref{fig:etaplot} for the standard case in which $\upsilon = m^2$. As we will see below, we can model both of these qualitative behaviors holographically by judiciously choosing the charge $Q$ of the black hole. However, there are also choices of $Q$ which yield dispersion relations that {\em can not} be captured by the simple equation~\eqref{eq:bulkplasmon}, but would require higher order corrections in $\omega$ and $k$ -- these are what we will refer to as `exotic' dispersion relations.
\begin{figure}[ht]
    \centering
    \begin{minipage}[t]{0.9\textwidth}
        \centering
        \includegraphics[width=0.7\linewidth]{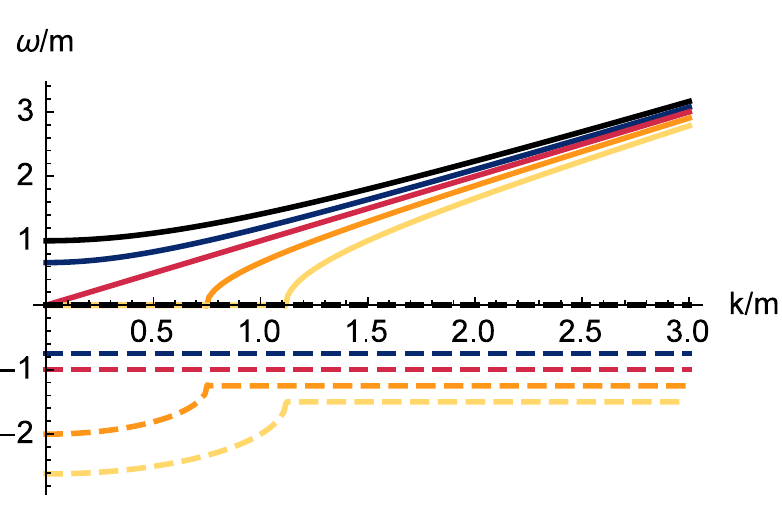}
        \caption{Dispersion relations from~\eqref{eq:bulkplasmon}, with $\upsilon = m^2$, for some under-, critically and over-damped systems. From top to bottom at $\eta/m=0,\,1.5,\,2,\,2.5,\,3$. Dashed curves are the imaginary parts.}\label{fig:etaplot}
    \end{minipage}
\end{figure}

The paper is organized as follows. In section~\ref{sec:plasmondisp} we present the dispersion relations for plasmon modes and explain how the seemingly complicated transitions, when varying the charge of the black hole, can be understood in terms of the pole structure of the dispersion relations. The analogous case for QNMs is presented in section~\ref{sec:QNMdisp}. In section~\ref{sec:uqd} we argue for a minimal extension of the toy model in (\ref{eq:bulkplasmon}) for strongly coupled processes based on our holographic results.
Finally, in section~\ref{sec:discussion} we summarize the results and discuss interesting future directions.

\section{Dispersion Relation for Plasmon Modes}
\label{sec:plasmondisp}

\begin{figure}[ht]
    \centering
    \begin{minipage}[t]{0.45\textwidth}
        \centering
        \includegraphics[width=1.0\linewidth]{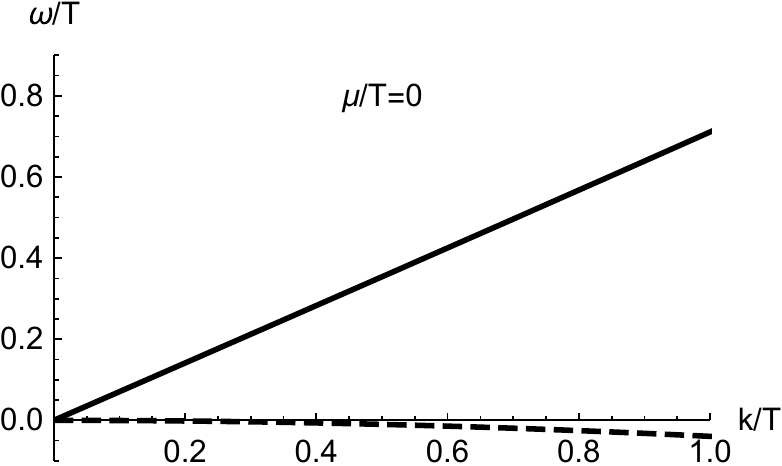}
        \caption{The dispersion relation for the lowest plasmon modes at $\mu/T=0$. The real part is chosen to be positive, although there also exists a mode with negative $\text{Re}[\omega/T]$. The imaginary part is the negative, dashed curve. It is worth pointing out that there also exists a purely diffusive mode very far below this one (at $\text{Im}[\omega]\approx-6$).}\label{fig:eps0}
    \end{minipage}\hfill
    \begin{minipage}[t]{0.45\textwidth}
        \centering
        \includegraphics[width=1.0\linewidth]{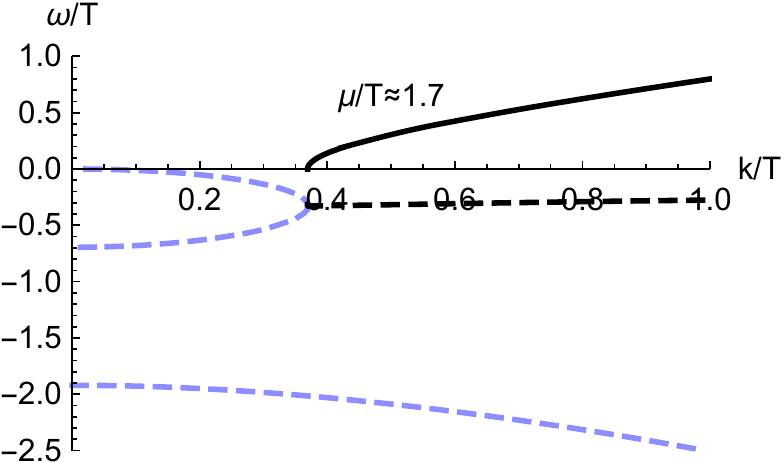}
        \caption{The lowest plasmon modes at $\mu/T\approx1.7$. Black lines are the real (solid) and imaginary (dashed) parts of a mode at the same $k/T$ (also technically two modes as there is one with a positive and one with a negative real part of $\omega$). Blue lines are purely imaginary. Note the third imaginary mode approaching from below.}\label{fig:eps40}
    \end{minipage}
\end{figure}

In this section we will investigate what happens if we start with an uncharged black hole, corresponding to $\mu/T = 0$ in the boundary QFT, leading to the linear zero sound dispersion relation as illustrated in figure~\ref{fig:eps0}, and gradually increase the chemical potential of the system. Due to scale invariance of the boundary CFT, there is only one relevant dimensionless parameter, $\mu/T$, and we can view the system as being held at a constant temperature $T$ while we change the chemical potential $\mu$.
For a weakly coupled system one would expect the zero sound mode to transform into the (bulk) plasmon mode according to the first three curves (red to black) in figure~\ref{fig:etaplot}. For a strongly coupled system instead, which holography generically gives rise to, we would expect the zero-sound mode to transform into the over-damped diffusive mode according to the last three curves (red to yellow) in figure~\ref{fig:eps0}. Note, however, that the route via the over-damped diffusive mode would present a slight conundrum since previous studies~\cite{Aronsson:2017dgf} revealed that for high values of $\mu/T$ the holographic plasmon dispersion is qualitatively well captured by the simple relation~\eqref{eq:bulkplasmon} for a {\em small} value of $\eta$, i.e.~by the standard quadratic and gapped dispersion. We will explain how this is resolved.

\begin{figure}[b]
    \centering
    \begin{minipage}[t]{0.45\textwidth}
        \centering
        \includegraphics[width=1.0\linewidth]{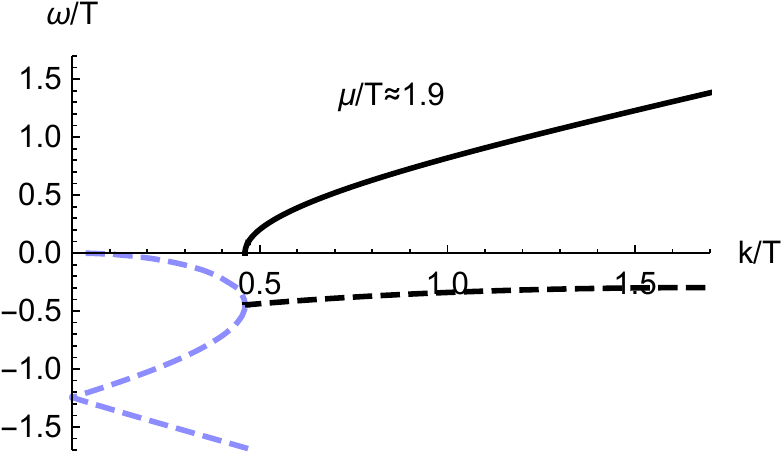}
        \caption{The lowest plasmon modes at $\mu/T\approx1.9$. Note here that the second and third purely diffusive mode meet.}\label{fig:eps43}
    \end{minipage}\hfill
    \begin{minipage}[t]{0.45\textwidth}
        \centering
        \includegraphics[width=1.0\linewidth]{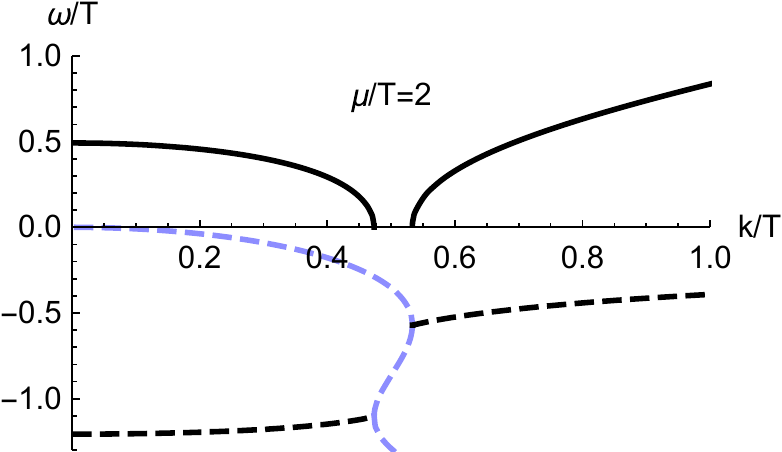}
        \caption{The lowest plasmon modes at $\mu/T=2$. The two diffusive modes that collided previously have now instead become propagating modes, and we see two bulges in the real part that approach each other.}\label{fig:eps46}
    \end{minipage}
\end{figure}

Turning on a chemical potential leads to the dispersion relation in figure~\ref{fig:eps40}, which is the over-damped diffusive mode as expected for a strongly coupled system.
When increasing $\mu/T$ further, the third diffusive mode in figure~\ref{fig:eps40} continues to move upwards and finally merges with the second diffusive mode at $k=0$, as illustrated in figure~\ref{fig:eps43}. The two diffusive modes combine into a propagating mode\footnote{Or rather two propagating modes, as there is a second one with the negative real part of the one plotted in the figures.}, c.f.~figure~\ref{fig:eps46}, which is the first example of the kind of `exotic' dispersion relation not captured by equation~\eqref{eq:bulkplasmon}.
Note in particular the damped intermediate region in $k/T$ for which there are no propagating modes\footnote{There will be higher propagating modes, but these decay very quickly and the three lowest modes we have exhibited are the ones dominating physical processes.}.

\begin{figure}[ht]
    \centering
    \begin{minipage}[t]{0.45\textwidth}
        \centering
        \includegraphics[width=1.0\linewidth]{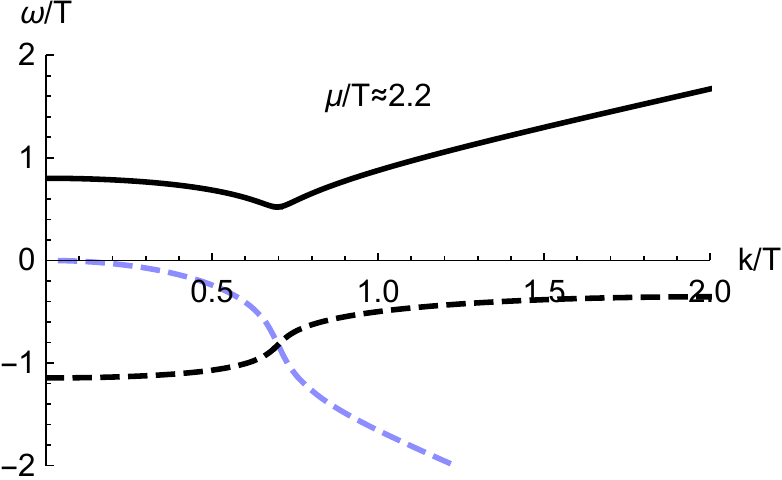}
        \caption{The lowest plasmon modes at $\mu/T\approx2.2$. Once the bulges in the real part meet, the previous three different diffusive mode becomes one, with a steep (negative) slope.}\label{fig:eps50}
    \end{minipage}\hfill
    \begin{minipage}[t]{0.45\textwidth}
        \centering
        \includegraphics[width=1.0\linewidth]{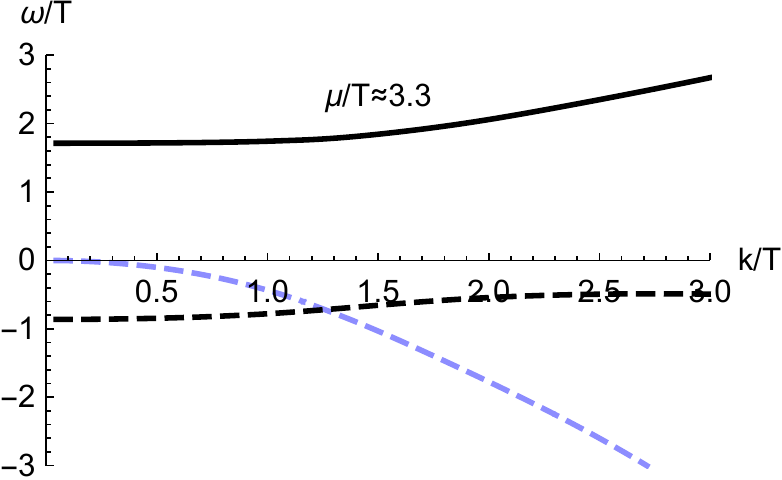}
        \caption{The lowest plasmon modes at $\mu/T=3.3$. Even at this relatively small $\mu/T$, we get something similar to the standard plasmon dispersion, and the agreement becomes increasingly better at larger $\mu/T$. }\label{fig:eps72}
    \end{minipage}
\end{figure}

Further turning up $\mu/T$ leads to a shrinking of the intermediate damped region, and eventually the two lowest intermediate diffusive poles will merge. The absence of propagating modes is now replaced by a local minimum in the dispersion relation for a specific value of $k/T$, which depends on $\mu/T$, as shown in figure~\ref{fig:eps50}. 

Going to even higher values of $\mu/T$ makes the local minimum in the dispersion relation disappear, and we have arrived at the same qualitative form of the dispersion relation as for the standard weakly coupled gapped quadratic relation.

\begin{figure}
    \centering
    \begin{minipage}[t]{0.9\textwidth}
        \centering
        \includegraphics[width=0.7\linewidth]{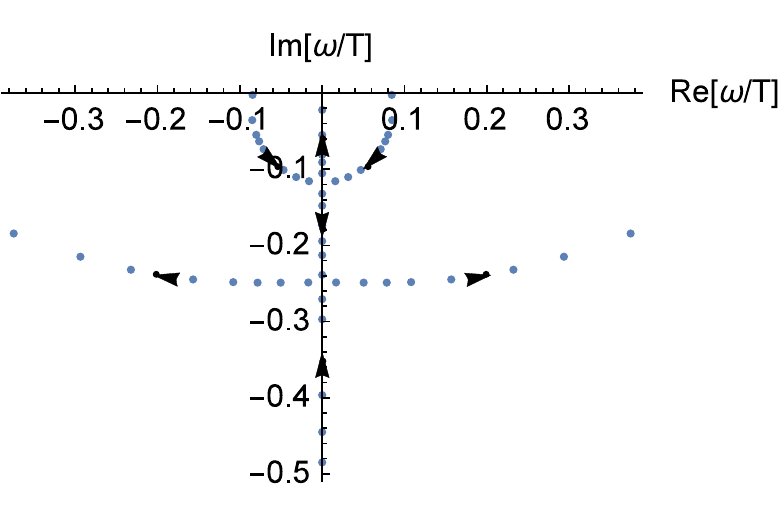}
        \caption{The movement of the three most dominant plasmon poles at $k/T=0.5$, as $\mu/T$ is increased.}\label{fig:modemerge_eps}
    \end{minipage}
\end{figure}

While this transition may seem strange at a first glance, from a conserved quantities point of view, it is one of the simplest possible solutions. The number of hydrodynamic (i.e.$~\omega(k\!\to\!0)\!\to\!0$) modes correspond to conserved currents, and should thus not change unless the system changes significantly (e.g.~at a phase transition or in the asymptotic regions $\mu/T=0$ or $\mu/T=\infty$). 
The longitudinal plasmonic sector for non-zero $\mu/T$ has only one hydrodynamic mode, the diffusive mode, and thus to continually turn into the hydrodynamic (zero) sound mode, this diffusive mode needs to be involved. This in turn is only possible if it collides with another (purely diffusive) mode, which can only happen (in a way related to the gapped mode that is) if the propagating plasmon mode (which is two modes, one with positive and one with negative $\text{Re}[\omega]$) turns into two diffusive modes and one of those collides with the hydrodynamic mode. This is exactly what we see in figures \ref{fig:eps0} to \ref{fig:eps72}, and it can also be visualized by the poles moving in the complex plane at a fix $k/T$, with increasing $\mu/T$, see figure \ref{fig:modemerge_eps}.

The range of $\mu/T$ where the exotic dispersion exists is relatively small, as demonstrated by figure~\ref{fig:eps43} and~\ref{fig:eps50}, just between 1.9 and 2.2. Thus, for this effect to be experimentally observed, the experiment would need to be very finely tuned. It is also worth noting that this transition occurs at $\mu\approx T$, which is not the typical range for experiments, as the chemical potential is typically of order eV and the temperature of order tenths of meV. That means, to be experimentally observed, the material needs to either be very hot (thousands of Kelvin) or very finely doped to be \emph{almost} neutral.

As an important consistency check, especially for the values of $\mu/T$ giving rise to the `exotic' dispersions, we have verified that the sum rule
\begin{equation}
    \lim_{k\to0}\int_0^\infty\frac{\mathrm{d}\omega}{\omega}\, \mathrm{Im}\,[\epsilon(\omega,k)^{-1}]=-\pi/2
\end{equation}
holds, as also verified in~\cite{Aronsson:2017dgf} for different values of $\mu/T$. 


\section{Dispersion Relation for QNMs}
\label{sec:QNMdisp}

Pioneering work in the study of QNMs for AdS-RN black holes can be found in~\cite{Wang:2000gsa,Berti:2003ud} for spherical BHs, and in~\cite{Edalati:2010pn, Davison:2011uk} for planar black holes, which is the case of relevance for application in condensed matter physics.

Following the analysis for the massive bulk plasmon mode in strongly coupled systems in the previous section, it is natural to examine the similar gapped quasinormal modes (QNMs), as arguably the same analysis holds also for these.  It is worth pointing out an important difference between them. While the diffusive mode is the only hydrodynamic (i.e.~$\omega(k\to0)\to0$) mode for plasmons, in addition there also exists a hydrodynamic sound mode for QNMs, which thus dominates at small energies and wave vectors.

\begin{figure}[ht]
    \centering
    \begin{minipage}[t]{0.45\textwidth}
        \centering
        \includegraphics[width=1.0\linewidth]{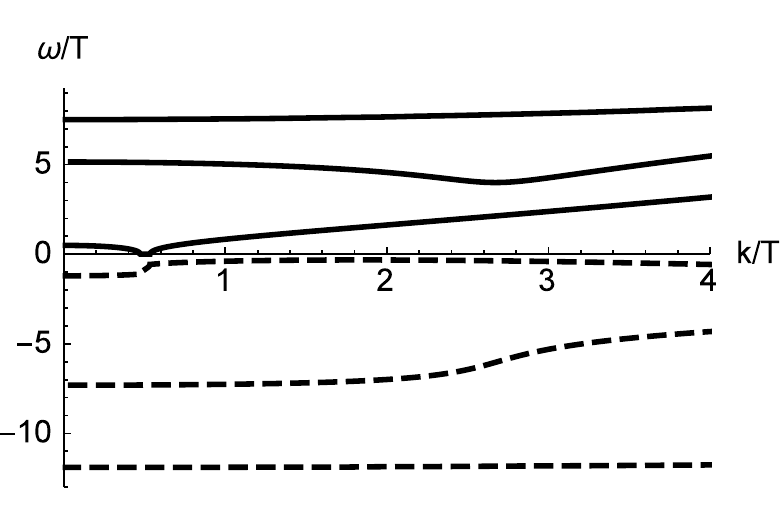}
        \caption{The three first gapped plasmon modes at $\mu/T=2$.}\label{fig:eps_gaps}
    \end{minipage}\hfill
    \begin{minipage}[t]{0.45\textwidth}
        \centering
        \includegraphics[width=1.0\linewidth]{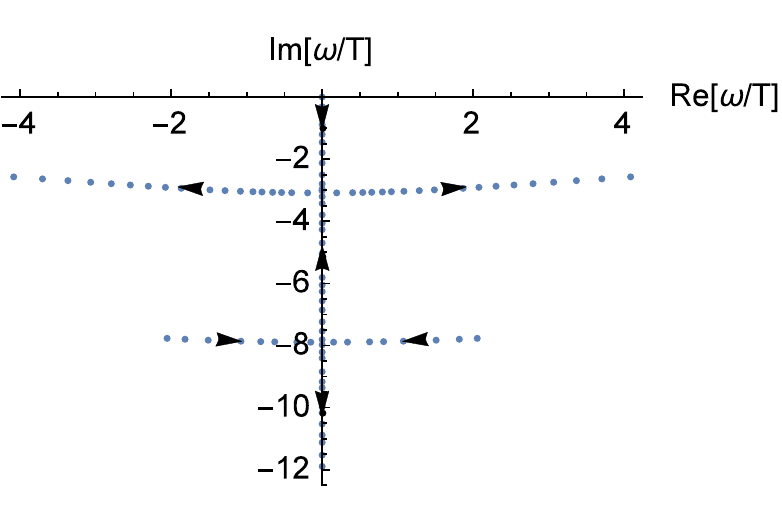}
        \caption{The movement of the QNM poles as $k/T$ is increased at $\mu/T\approx1.3$. 
        }
        \label{fig:modemerge}
    \end{minipage}
\end{figure}

Thus, we focus mostly on the next to leading mode, where we indeed find a similar behavior as for plasmons, which we illustrate in figures~\ref{fig:qnm0}-\ref{fig:qnm100} below. Higher modes are for completeness illustrated in figure~\ref{fig:eps_gaps}.

\begin{figure}[ht]
    \centering
    \begin{minipage}[t]{0.45\textwidth}
        \centering
        \includegraphics[width=1.0\linewidth]{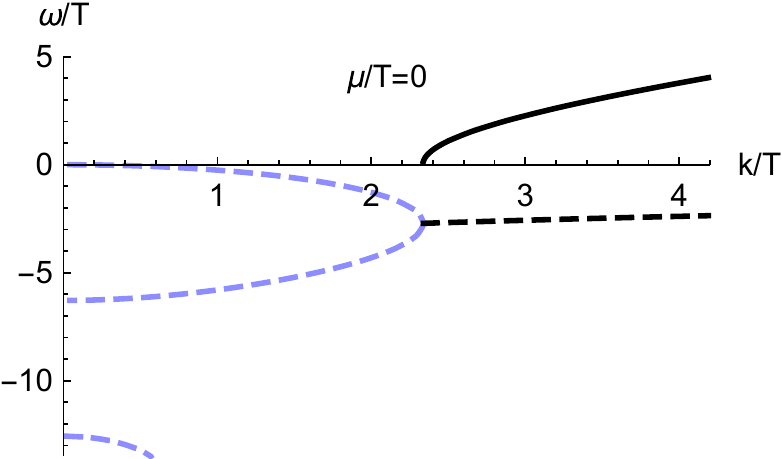}
        \caption{The dispersion relation for the lowest QNMs at $\mu/T=0$. Solid lines are real parts and dashed lines are imaginary parts. Blue lines are purely imaginary. The real part is chosen to be positive, although there also exists a mode with negative $\text{Re}[\omega/T]$. For clarity, we have omitted the ever present sound mode that is also hydrodynamical.}\label{fig:qnm0}
    \end{minipage}\hfill
    \begin{minipage}[t]{0.45\textwidth}
        \centering
        \includegraphics[width=1.0\linewidth]{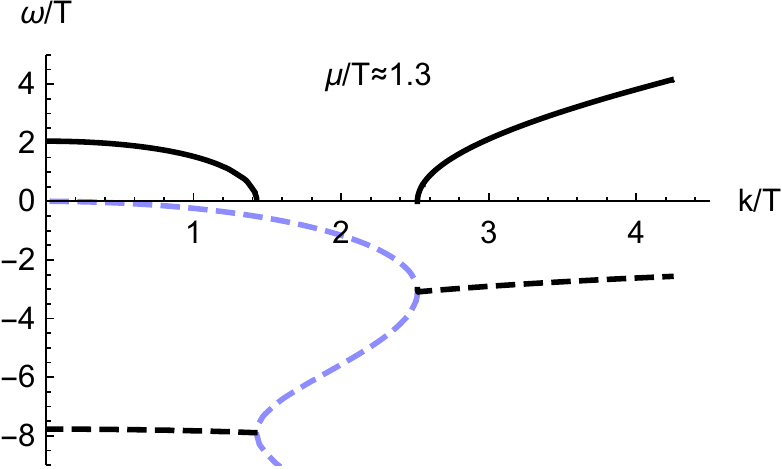}
        \caption{The lowest QNMs at $\mu/T\approx1.3$. Note here that the second and third purely diffusive mode in figure~\ref{fig:qnm0} have met and given rise to propagating modes for small $k/T$.}\label{fig:qnm30}
    \end{minipage}
\end{figure}


\begin{figure}[ht]
    \centering
    \begin{minipage}[t]{0.45\textwidth}
        \centering
        \includegraphics[width=1.0\linewidth]{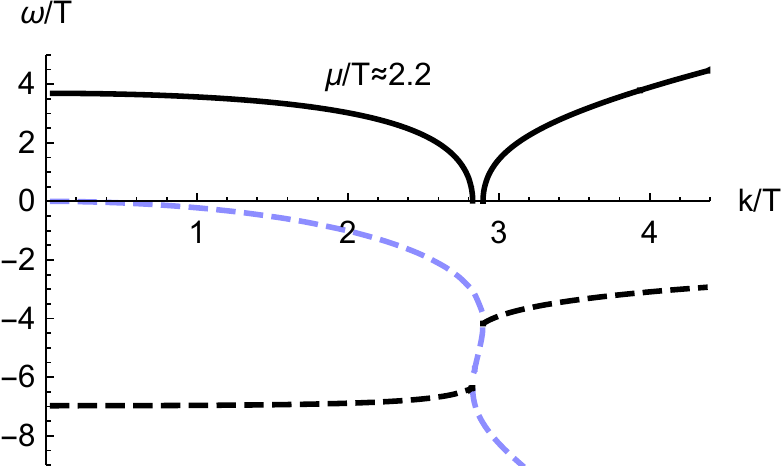}
        \caption{The lowest QNMs at $\mu/T\approx2.2$. The two bulges in the real part are approaching each other, and the diffusive modes start to resemble a single mode with a steep slope.}\label{fig:qnm50}
    \end{minipage}\hfill
    \begin{minipage}[t]{0.45\textwidth}
        \centering
        \includegraphics[width=1.0\linewidth]{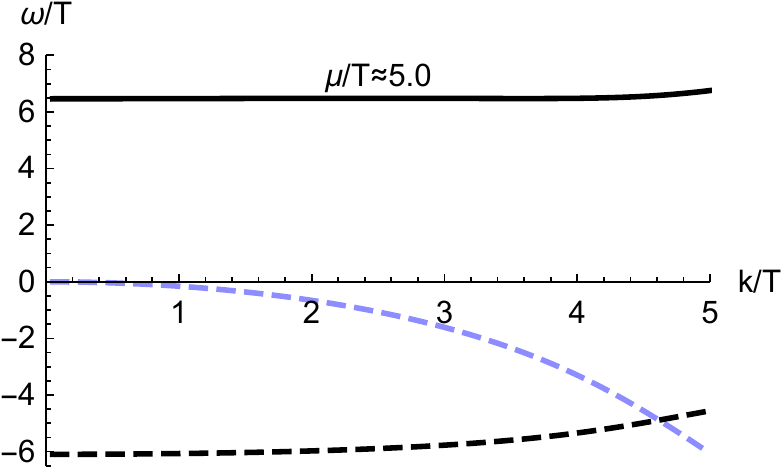}
        \caption{The lowest QNMs at $\mu/T=5.0$. Even at this relatively small $\mu/T$, we get something similar to the expected gapped dispersion and the agreement becomes increasingly better at larger $\mu/T$.}\label{fig:qnm100}
    \end{minipage}
\end{figure}


In figure~\ref{fig:qnm0} the dispersion relation for the lowest modes are shown, at $\mu/T=0$. To make the figures less cluttered, the sound mode, which is essentially unchanged when tuning $\mu/T$, has been omitted (c.f. figure~\ref{fig:eps0}). When turning up $\mu/T$, the second and third mode approaches each other, before meeting and giving rise to a new propagating mode, shown in figure~\ref{fig:qnm30}. This new bulge expands further as $\mu/T$ is increased until the two propagating modes meet and merge, c.f.~figure~\ref{fig:qnm50}, yielding a local minimum. Further increasing $\mu/T$ shrinks the local minimum near the merging point, until ultimately, a more familiar dispersion relation is obtained for larger $\mu/T$, figure~\ref{fig:qnm100}. 

Another way of presenting the exotic region, is by the movement of the poles as $k/T$ is increased, shown in figure~\ref{fig:modemerge}. 

A minor difference for QNMs compared to plasmons is that the interval in $\mu/T$ at which this exotic behavior is demonstrated is slightly larger for QNMs than for plasmons.

\section{Unconventional Quantum Damping}
\label{sec:uqd}

Taking a closer look at the dispersions in sections~\ref{sec:plasmondisp} and~\ref{sec:QNMdisp} reveals that the damping effects of modes in strongly correlated systems have a much richer structure than can be captured in the toy model~\eqref{eq:bulkplasmon} we have introduced above, inspired by the Caldeira--Leggett approach to quantum damping. In particular the behavior of poles, as shown in figure~\ref{fig:modemerge_eps}, respectively~\ref{fig:modemerge}, illustrates rather clearly that the damping due to the interaction of the plasmon, or QNM, with other effects in the system, can not adequately be captured by a simple linear damping term~$i\eta\omega$. Such effects are not unprecedented, and have occasionally been referred to as the `unzipping' of poles on the imaginary axis~\cite{WitczakKrempa:2013ht}. In the case of QNMs, however, such effects have usually only been noticed in subleading modes, due to the presence of the sound mode omitted in the figures, such that this `exotic' behavior would only result in relatively minor, secondary corrections needing to be included. What makes the situation peculiar for the plasmon mode is that this contribution is dominant, meaning that an improved understanding of quantum damping effects is indisputably necessary to fully understand the underlying physical processes.

While a full investigation of this phenomenon will require much more input, holography can provide a first direction in which research can be pursued. Based on the results presented here, a minimal extension of the basic model~\eqref{eq:bulkplasmon} can be argued for as follows.
The main contributors to the unconventional dispersion relation are the three most dominant poles, where two of them can unzip and then zip again on the imaginary axis. Keeping the dynamics of the model as simple as possible, i.e.~by not rearranging the kinematic term, but by just allowing for a slightly more intricate way of interactions due to a potential and coupling to the `heat bath', it is rather straightforward to see that the major contribution in the dispersion can be captured with the simple ansatz
\begin{equation}
    \label{eq:improved_qd}
    \omega^2 - c^2 k^2 - \frac{i\,\eta\,\omega - \upsilon}{i\,\eta\,\omega - \upsilon_\infty}\upsilon_\infty = 0 \,.
\end{equation}
For small $\omega$ the behavior is almost indistinguishable from the Caldeira--Leggett inspired model. The intriguing difference appears at large $\omega$, where the effect of quantum damping results in the mode experiencing an `effective' potential $\upsilon_\infty$ instead of linear damping.

We will leave it for future research to investigate how such unconventional quantum damping effects are realized in condensed matter systems.

\section{Discussion}
\label{sec:discussion}


In this paper we present a thorough investigation of plasmon modes in the RN-metal, a holographic dual for a strongly interacting phase of matter, which builds on previous work on how to model plasmonic excitations holographically, relevant for strongly correlated systems~\cite{Aronsson:2017dgf, Aronsson:2018yhi}. We also compare the behavior of the plasmon modes to that of QNMs. A peculiar feature we uncover is an intermediate region where the dispersion is over-damped. While an upper limit to a damped region is to be expected -- when the momentum $k$ becomes large enough, interaction with the system will be negligible -- having a lower limit is a clear sign that processes inside the system must be highly non-trivial as a result of strongly correlated physics. As such, first order damping effects one would expect from the usual Caldeira--Leggett approach are clearly not sufficient to describe the underlying processes. Based on our results, we propose a minimal generalization of the basic model for quantum damping~\eqref{eq:bulkplasmon} to identify the higher corrections necessary to model the `exotic' dispersion relations.


We also wish to emphasize that the intermediate over-damped region is a dominant effect in the plasmon dispersion, while, it is only sub-leading in the case of QNMs due to presence of the sound mode. This is particularly relevant with regard to the currently growing interest in the field of plasmonics. As mentioned in section~\ref{sec:plasmondisp}, the bandwidth in which the modes are over-damped could potentially be adjusted with a carefully fine-tuned choice of doping, opening up the possibility to employ strongly correlated materials in the conception of band-pass filters and similar devices for miniaturized circuits using plasmonic excitations.


While the RN-metal, due to its simplicity, is very useful to illustrate certain concepts from the holographic approach, it has the drawback that is it rather limited in free parameters. Thus, it would be desirable to apply the holographic plasmon formalism~\cite{Aronsson:2017dgf, Aronsson:2018yhi} to a more sophisticated set of backgrounds. One such candidate is the `electron star'~\cite{Hartnoll:2010gu}, or its finite temperature version~\cite{Puletti:2010de,Hartnoll:2010ik}, sometimes referred to as the `electron cloud'. An interesting issue to investigate is whether having just one dimensionless parameter in the RN model forces the dispersion to take the route via exotic dispersion relations, and if a different route can be taken when more parameters are available. This is work in preparation~\cite{ECpaper}.

\acknowledgments

We would like to thank Richard Davison, Andreas Isaksson, Napat Poovuttikul and Henk Stoof for valuable discussions and useful comments during the process of writing this paper. This work is supported by the Swedish Research Council.




\bigskip

\bibliographystyle{JHEP}
\bibliography{ExoticDispersion}

\end{document}